\documentclass[aps,prl,twocolumn,superscriptaddress]{revtex4}
\usepackage{amsmath}
\usepackage{epsfig}
\usepackage{rotating,dcolumn}
\newcommand\BiFe{Bi$_{2}$Fe$_{4}$O$_{9}$}

\begin{document}

\title{Magnetic frustration in an iron based Cairo pentagonal lattice}

\author{E. Ressouche}
\email[]{ressouche@ill.fr}
\affiliation{CEA/Grenoble, INAC/SPSMS-MDN, 17 rue des Martyrs, 38054 Grenoble Cedex 9, France}
\author{V. Simonet} 
\affiliation{Institut N\'eel, CNRS \& Universit\'e Joseph Fourier, BP166, 38042 Grenoble Cedex 9, France}
\author{B. Canals} 
\affiliation{Institut N\'eel, CNRS \& Universit\'e Joseph Fourier, BP166, 38042 Grenoble Cedex 9, France}
\author{M. Gospodinov} 
\affiliation{Institute of Solid State Physics, Bulgarian Academy of Sciences, 1184 Sofia, Bulgaria}
\author{V. Skumryev} 
\affiliation{Instituci\'o Catalana de Recerca i Estudis Avanats (ICREA) and Departament de F\'{\i}sica, Universitat Aut\`onoma de Barcelona, 08193 Bellaterra, Spain}
\date{\today}

\begin{abstract}
The Fe$^{3+}$ lattice in the \BiFe~compound is found to materialize the first analogue of a magnetic pentagonal lattice. Due to its odd number of bonds per elemental brick, this lattice, subject to first neighbor antiferromagnetic interactions, is prone to geometric frustration. The \BiFe~magnetic properties have been investigated by macroscopic magnetic measurements and neutron diffraction. The observed non-collinear magnetic arrangement is related to the one stabilized on a perfect tiling as obtained from a mean field analysis with  direct space magnetic configurations calculations. The peculiarity of this structure arises from the complex connectivity of the pentagonal lattice, a novel feature compared to the well-known case of triangle-based lattices.
\end{abstract}

\maketitle

\begin{figure}[h]
\includegraphics[width=8.5cm]{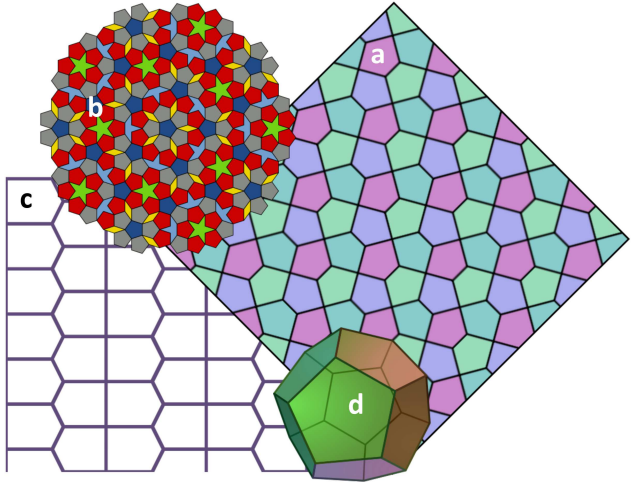}
\caption{(color online) Pentagonal lattices.  (a) Cairo pentagonal tiling, (b) Penrose pentagonal tiling,  (c) pentagonal lattice studied in refs. \onlinecite{waldor, bhaumik, moessner}, (d) dodecahedron.}
\label{pentagon}
\end{figure}

The pentagon, a 5-edges polygon, is an old issue in mathematical recreation. It forms the faces of the dodecahedron, one of the platonic solids whose shape is reproduced in biological viruses and in some metallic clusters~\cite{kong}. The main peculiarity of this polygon is that, contrary to triangles, squares or hexagons, it is impossible to tile a plane with congruent {\bf regular} pentagons, the tilings must involve additional shapes to fill the gaps~\cite{Kepler}, like in the Penrose lattice (Fig.~\ref{pentagon}). It exists however several possibilities of tessellation of a plane with {\bf non-regular} pentagons, a famous one being the Cairo tessellation whose name was given because it appears in the streets of Cairo and in many Islamic decorations (Fig.~\ref{pentagon}). 

Such a lattice could attract interest in the field of geometric frustrated magnetism.  Indeed, frustration usually arises when all pairs of magnetic interactions are not simultaneously satisfied in a system due to the lattice topology, resulting in unusual physical properties \cite{frustration}. Today, most of the studied systems are mainly  based on triangles (or tetrahedra in three dimensions), and can be made perfect from both structural (regular polygons) and magnetic (equal first neighbor magnetic interactions) points of view. This is not the case with pentagonal lattices that must involve only non-regular pentagons to fill the space. The experimental realization of such a model system is thus a unique opportunity that could bring new interesting features in this field of magnetic frustration. In this letter we show that the  Fe$^{3+}$ lattice in \BiFe~materializes the first analogue of a magnetic pentagonal lattice, and we interpret its peculiar non-collinear magnetic arrangement with respect to magnetic frustration.

\begin{figure*}[t]
\begin{minipage}[t]{.3\linewidth}
\includegraphics[width=4.8cm]{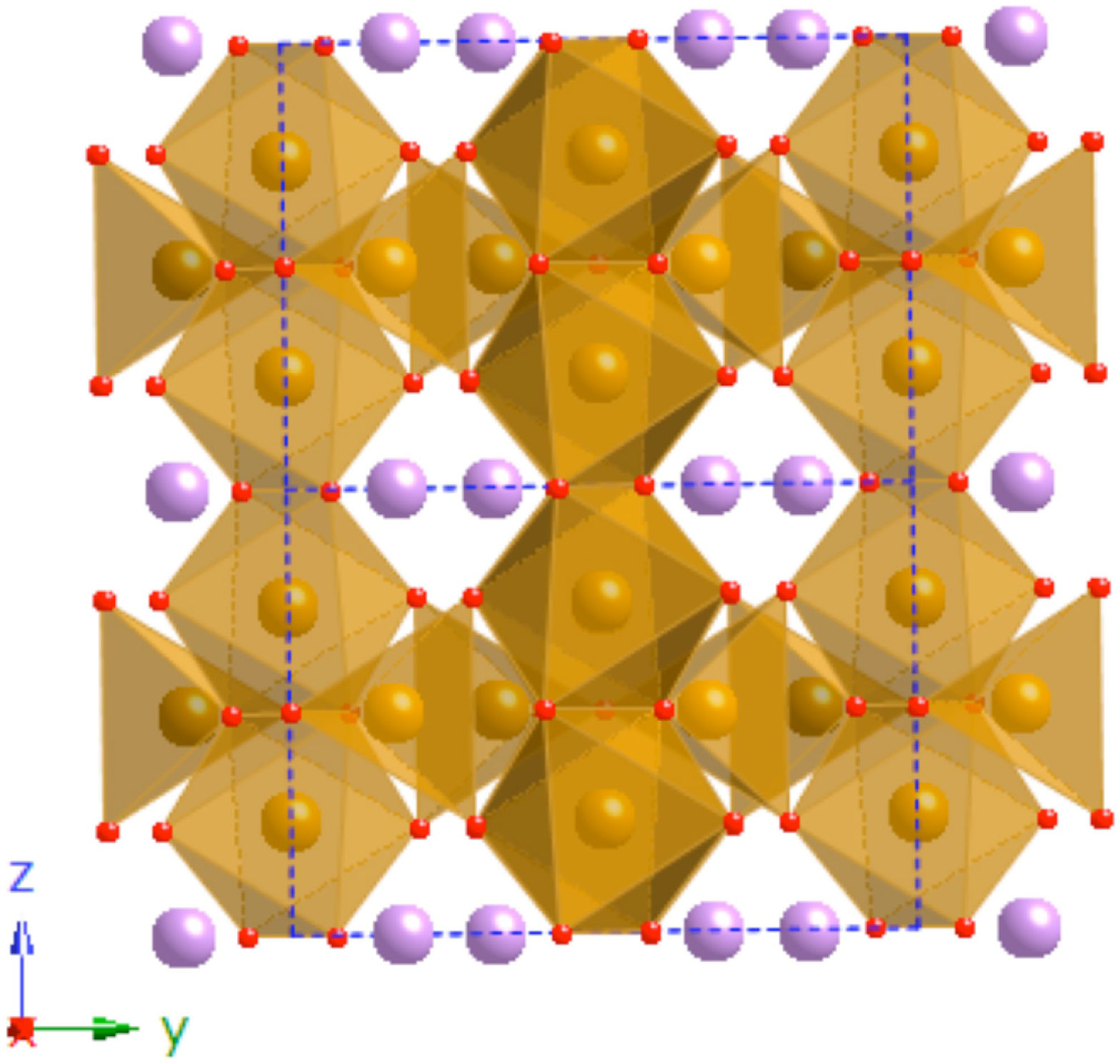}
\end{minipage}
\begin{minipage}[t]{.3\linewidth}
\includegraphics[width=5.8cm]{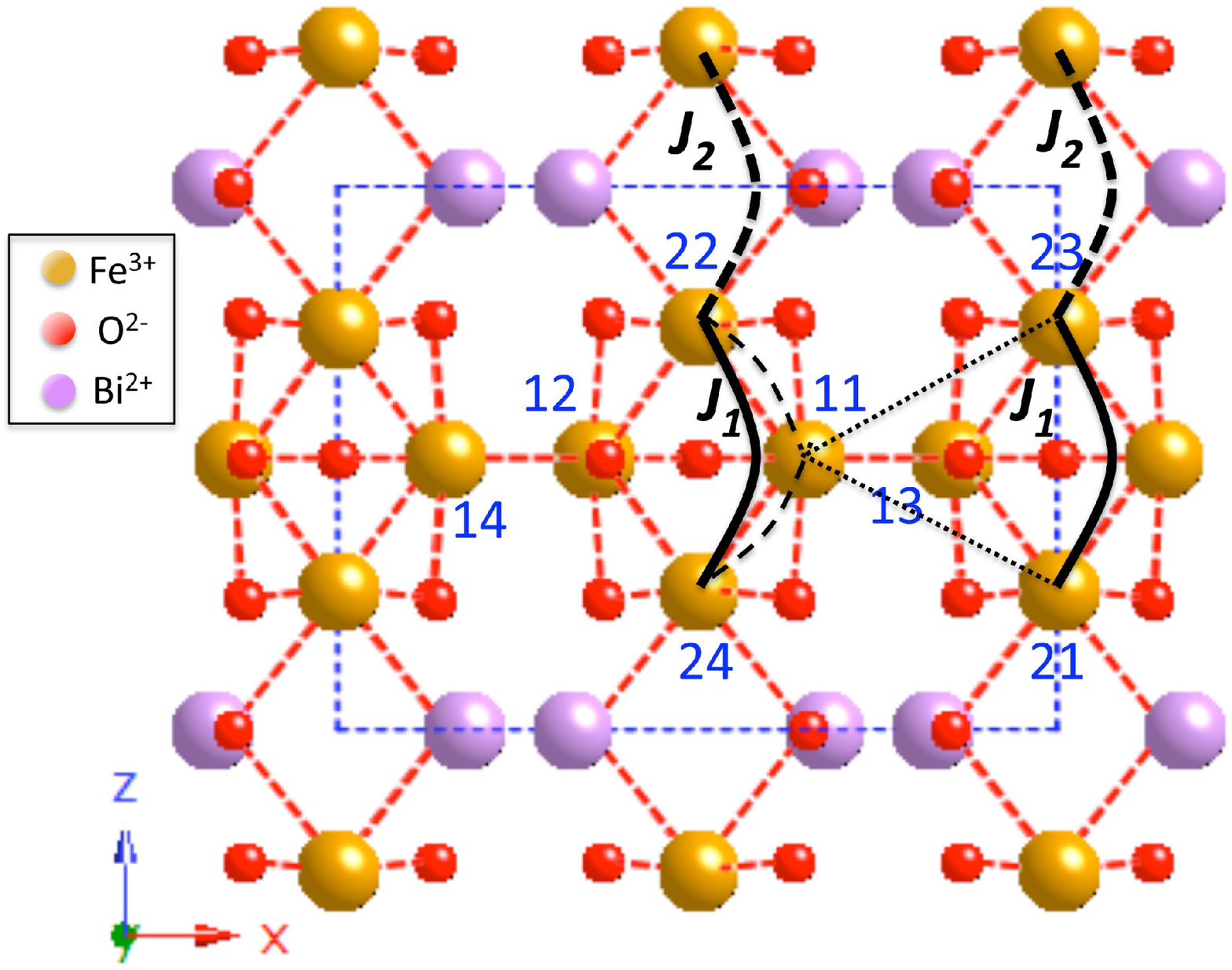}
\end{minipage}
\begin{minipage}[t]{.3\linewidth}
\includegraphics[width=4.6cm]{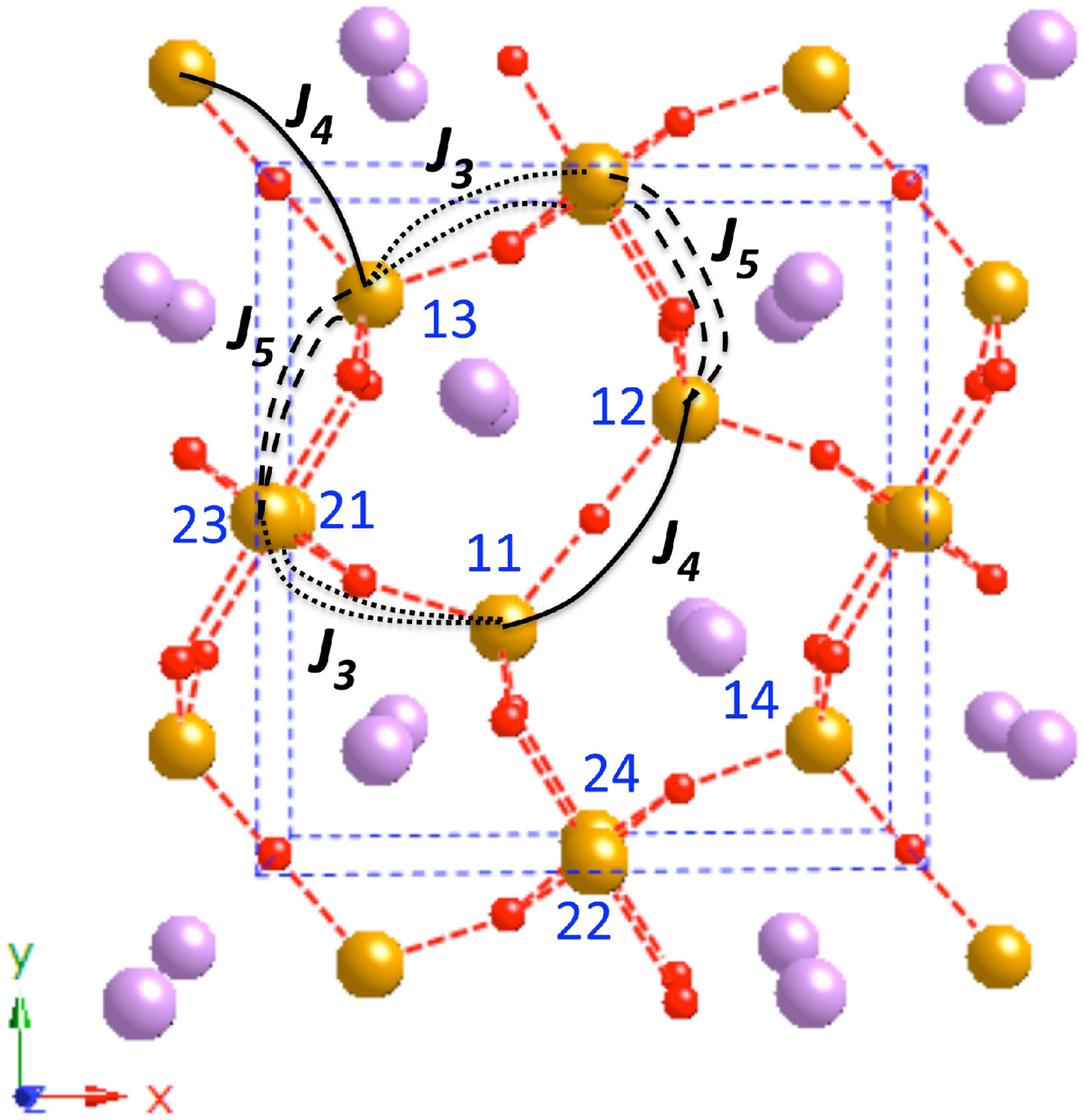}
\end{minipage}
\caption{(color online) The \BiFe\ compound (space group {\it Pbam}; a = 7.965(4), b = 8.448(5), c = 6.007(3) \AA\ at 300 K). (left): View of the ~unit cell with the environment of the two Fe ions. Bi atoms are purple, Fe yellow and O red. Projection of the structure on the ($a$, $c$) plane (middle) and on the ($a$, $b$) plane (right). The atoms labels are in blue. Black lines materialize the magnetic exchange interactions.}
\label{structure}
\end{figure*}

 \BiFe\ is a common by-product in the synthesis of the multiferroic compound BiFeO$_3$, and has been claimed recently to display itself multiferroic properties \cite{singh}. This was the initial motivation of our study: this finding requires a precise determination of the symmetries in the magnetic ordered phase, since ferroelectricity can only arise in polar space groups. Such a precise determination has never been carried out on single crystals of  \BiFe, and only powder neutron diffraction measurements have been reported \cite{Shamir} with no definitive conclusions due to the complexity of this orthorhombic structure. Actually, there are two different sites of four iron atoms in \BiFe: Fe$_1$ occupies a tetrahedral position and Fe$_2$ an octahedral one. In the structure, columns of edge-sharing Fe$_2$ octahedra form chains along  the {\bf c} axis, and these chains are linked together by corner-sharing Fe$_1$ tetrahedra and Bi atoms (Fig.~\ref{structure}). As a result, the lattice formed by the different Fe$^{3+}$ magnetic atoms is quite remarkable. From a simple examination of the structure, five main magnetic superexchange interactions, $J_1$ to $J_5$, can be identified (Fig.~\ref{structure}). The structure can be simply viewed as layers perpendicular to the {\bf c} axis: the magnetic coupling of these layers along the {\bf c} direction involves only Fe$_2$ atoms in the octahedron columns and is achieved through two interactions, $J_1$ between the atoms within the unit cell and $J_2$ between the atoms of adjacent cells. Within a layer, each Fe$_1$ interacts with its nearest neighbour Fe$_1$ via $J_4$ and with two nearest neighbour Fe$_2$  pair (located above and below the mean plane) via $J_3$ and $J_5$. The projection of this layer along {\bf c} forms a pentagonal lattice with three slightly different bonds per pentagon (Fig.~\ref{structure}). This geometry is equivalent to a distorted Cairo lattice, the perfect one involving convex equilateral pentagons with equal-length sides, but with different associated angles (Fig.~\ref{pentagon}).

\begin{figure}[h]
\includegraphics[width=9cm]{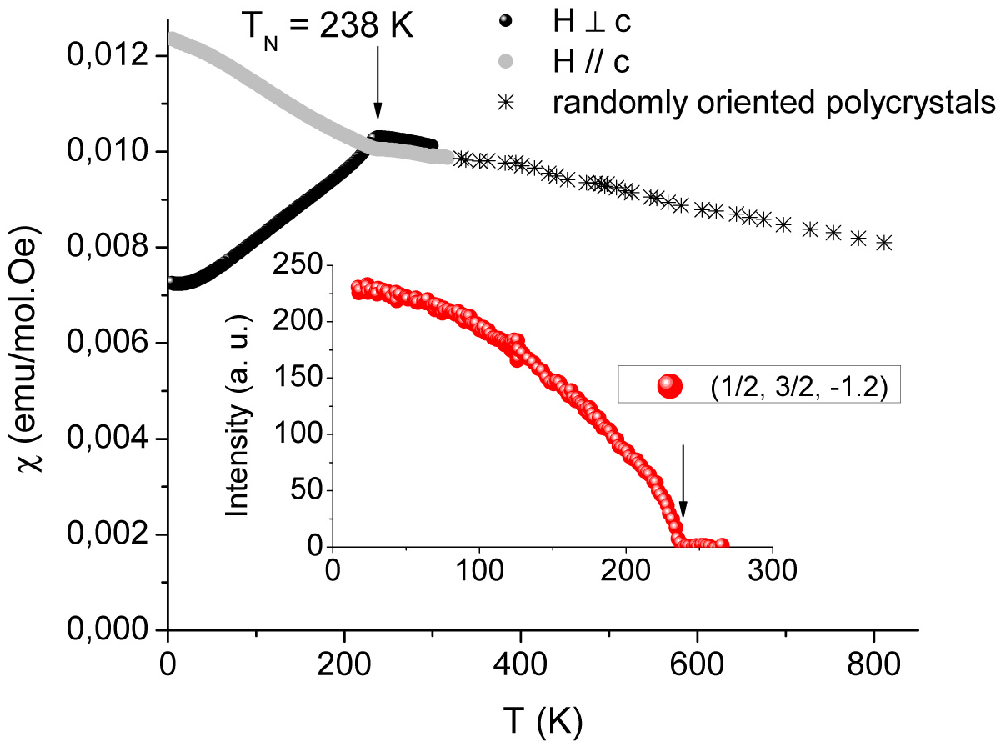}
\includegraphics[width=7cm]{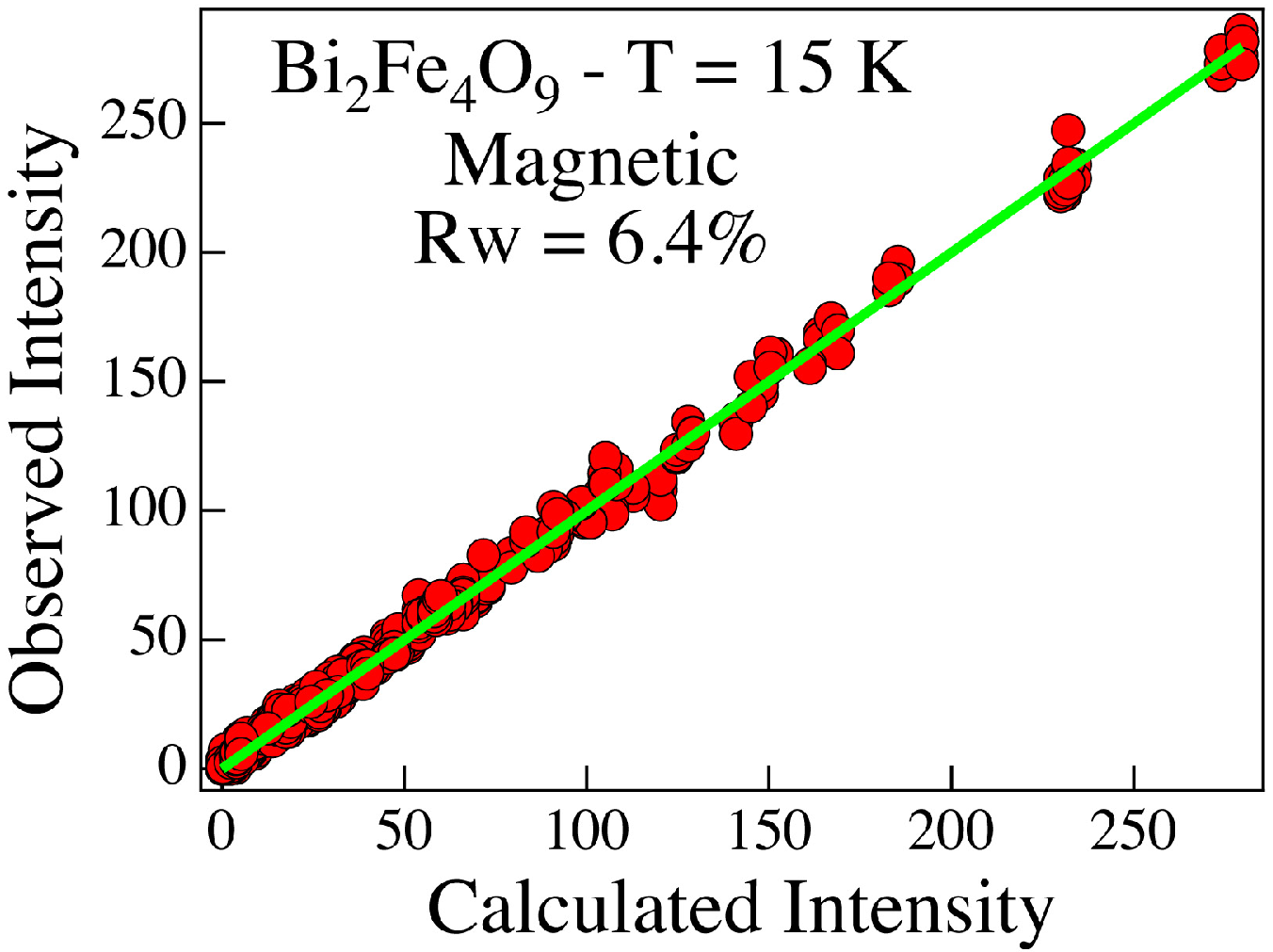}
\caption{(color online) Linear magnetic susceptibility of \BiFe\ as a function of temperature measured in magnetic fields of 0.1 T and 0.7 T below and above 300 K respectively. Inset: Temperature variation of the $(1/2, 3/2, -1/2)$ magnetic Bragg peak in zero field from neutron diffraction. \\ (Bottom) Graphical representation of the magnetic structure refinement at T = 15 K in \BiFe~: observed versus calculated intensities.}
\label{Magn}
\end{figure}

What should be expected from such a geometry and what is actually observed ? Up to now, no experimental studies and very few theoretical calculations are available in the literature \cite{waldor, bhaumik, moessner}.  They only concern a very particular pentagonal tiling, derived from the hexagonal lattice (see lattice {\bf c} of Fig.\ref{pentagon}). Calculations within the antiferromagnetic Ising model on this lattice yields a disordered ground state with a finite entropy per spin  \cite{waldor, moessner}, whereas  in the Heisenberg model  the classical ground state of the perfect pentagonal lattice has not yet been solved \cite{bhaumik}.  \\
Single crystals of \BiFe\ were grown by the high temperature solution growth method using a flux of Bi$_2$O$_3$ \cite{Ressouche}. The temperature and magnetic field dependences of the magnetization along the principal crystallographic axes were measured using a SQUID magnetometer (Quantum Design) in the temperature range 2-380 K and in fields up to 5.5 T. The reported data, measured on the very same crystal used for the neutron diffraction study (2.5 x 2 x 1.5 mm$^3$), were corrected for the demagnetizing field effect. The DC susceptibility between 300 K and 800 K was measured on a home made Faraday-type magnetic balance in a magnetic field of 0.7 T,  using an assembly of several randomly oriented single crystals.  The reported susceptibility was corrected from diamagnetism using the Pascal's constants $\chi_{\rm dia}$ = -198 10$^{-6}$ emu/mol. 
Above room temperature, the magnetic susceptibility of  \BiFe\ is found to obey a Curie-Weiss law with an extrapolated paramagnetic temperature $\theta_p\approx$ - 1670 K and an effective magnetic moment $\mu_{\rm eff}=$ 6.3(3) $\mu_B$ per iron atom (see Fig.~\ref{Magn}). This later value is compatible with the 5.9 $\mu_B$ expected for Fe$^{3+}$ ions ($J$=$S$=5/2). When decreasing the temperature,  an evident drop in the ({\bf a},{\bf b}) plane susceptibility and  a well pronounced increase of the {\bf c} axis susceptibility are observed at T$_N = 238(2)$~K,  indicating a transition toward a magnetic long range order. This N\'eel temperature is in agreement with  the results on single crystal reported in ref.~\onlinecite{giaquinta} but some 20 K lower that the one reported in refs.~\onlinecite{singh, Shamir, tutov} for sintered samples. The observed ratio between the N\'eel temperature and the paramagnetic one,  $\theta_p/$T$_N\approx$~7, suggests, as expected from the geometry, a large degree of frustration.

Further insights concerning the magnetic arrangement below T$_N$ were obtained using neutron diffraction. The neutron diffraction experiments were performed on the two CEA-CRG diffractometers  D23 and D15 at the Institut Laue-Langevin (Grenoble, France) using wavelengths of respectively  $\lambda = 1.280$~\AA\ and $1.173$\AA\ . Complete data collections have been made at two different temperatures $T= 15$~K and $T = 300$~K. The experimental data were corrected from absorption using the CCSL library \cite{Brown93}. The structural and magnetic arrangements were refined using the least-square programs MXD \cite{Wolfers90} and FULLPROF \cite{Juan92}, including extinction corrections \cite{Becker74a}.

\begin{table}[h]
\begin{center}
\begin{tabular}{lcccccccc}
\hline
                           &                &             &       &                              &&                                  && \\
			       & x & y & z &	$\mu$        & &     $\theta$                                   & &	$\phi$          \\ 
\hline
                           &                &             &       &                              &&                                  && \\
Fe${_1}^{1}$   & 0.3540 & 0.3361 & 0.5 &      3.52(1)      	& &           100.1(1)    	   & &         90      \\ 
Fe${_1}^{2}$   & 0.6460 & 0.6639 & 0.5 &      3.52(1)      	& &      $\theta_1^1$ + 180  & &         90      \\ 
Fe${_1}^{3}$   & 0.1460 & 0.8361 & 0.5 &      3.52(1)      	& &     $\theta_1^1$ - 90      & &          90      \\ 
Fe${_1}^{4}$   & 0.8540 & 0.1639 & 0.5 &      3.52(1)      	& &      $\theta_1^1$ + 90    & &         90     \\ 
\hline
                           &                &             &       &                              &&                                  && \\
Fe${_2}^{1}$   & 0 & 0.5& 0.2582       &      3.73(1)      & &        -105.4(1)    	          & &         90      \\
Fe${_2}^{2}$    & 0 & 0.5& 0.7418      &      3.73(1)      & &      $\theta_2^1$ 		   & &         90     \\ 
Fe${_2}^{3}$   & 0.5 & 0& 0.7418       &      3.73(1)      & &      $\theta_2^1$ +90        & &         90     \\  
Fe${_2}^{4}$    & 0.5 & 0& 0.2582      &      3.73(1)      & &      $\theta_2^1$ + 90       & &         90      \\ 
\hline
\end{tabular}
\end{center}
\caption{Coordinates of the Fe$^{3+}$ ions in  \BiFe\ and spherical components of their magnetic moments in the first magnetic domain. The magnetic moments are defined in spherical coordinates by $(\mu, \theta, \phi)$. In cartesian coordinates (X,Y,Z) where X is along {\bf a}, Y along {\bf b}  and Z along {\bf c}, the magnetic moment components are : $ M_X = \mu\ cos\theta\ sin\phi, M_Y = \mu\ sin\theta\ sin\phi, M_Z = \mu\ cos\phi$.}
\label{resneut}
\end{table}

\begin{figure}[h]
\includegraphics[width=6cm]{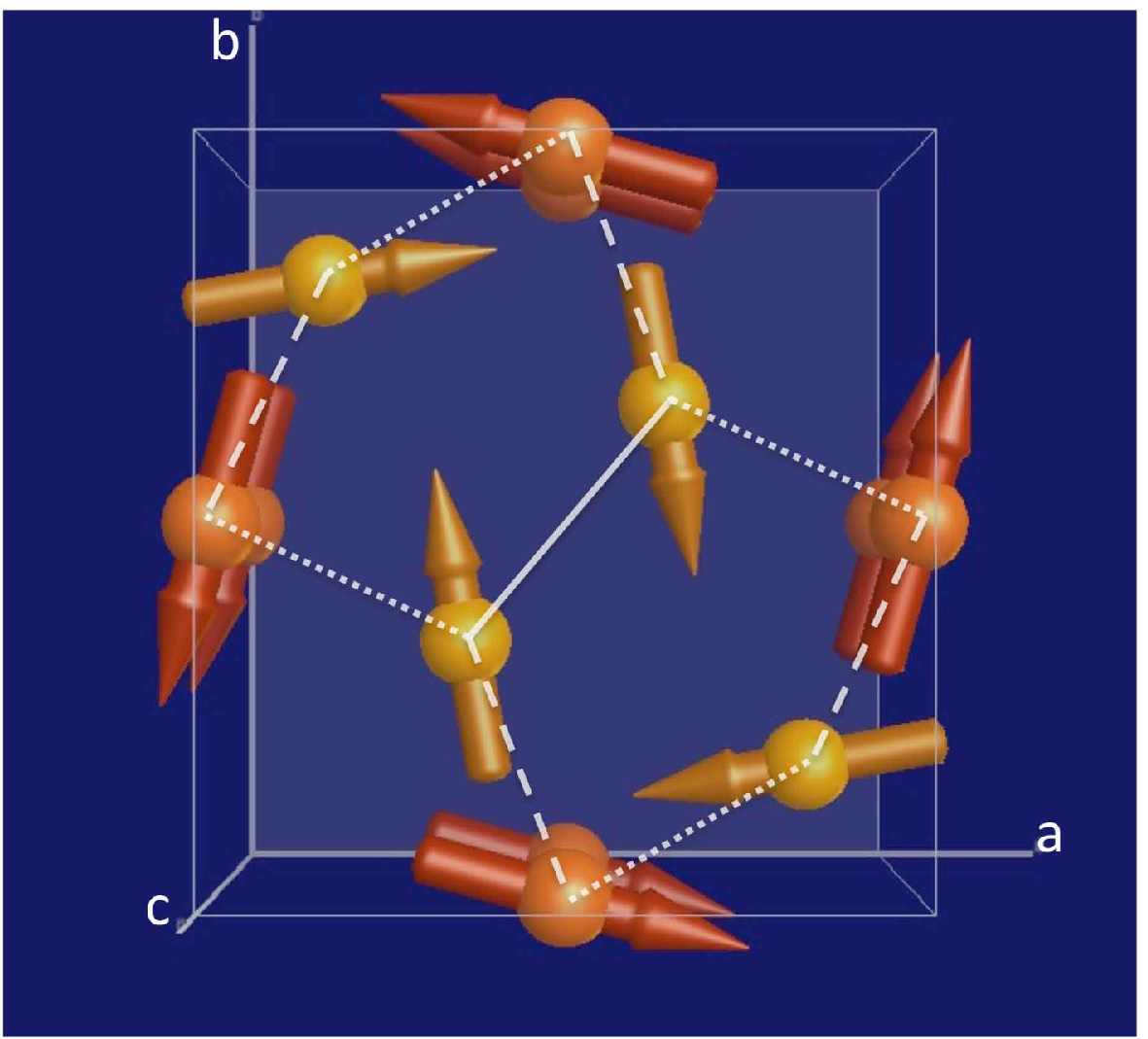}
\includegraphics[width=6cm]{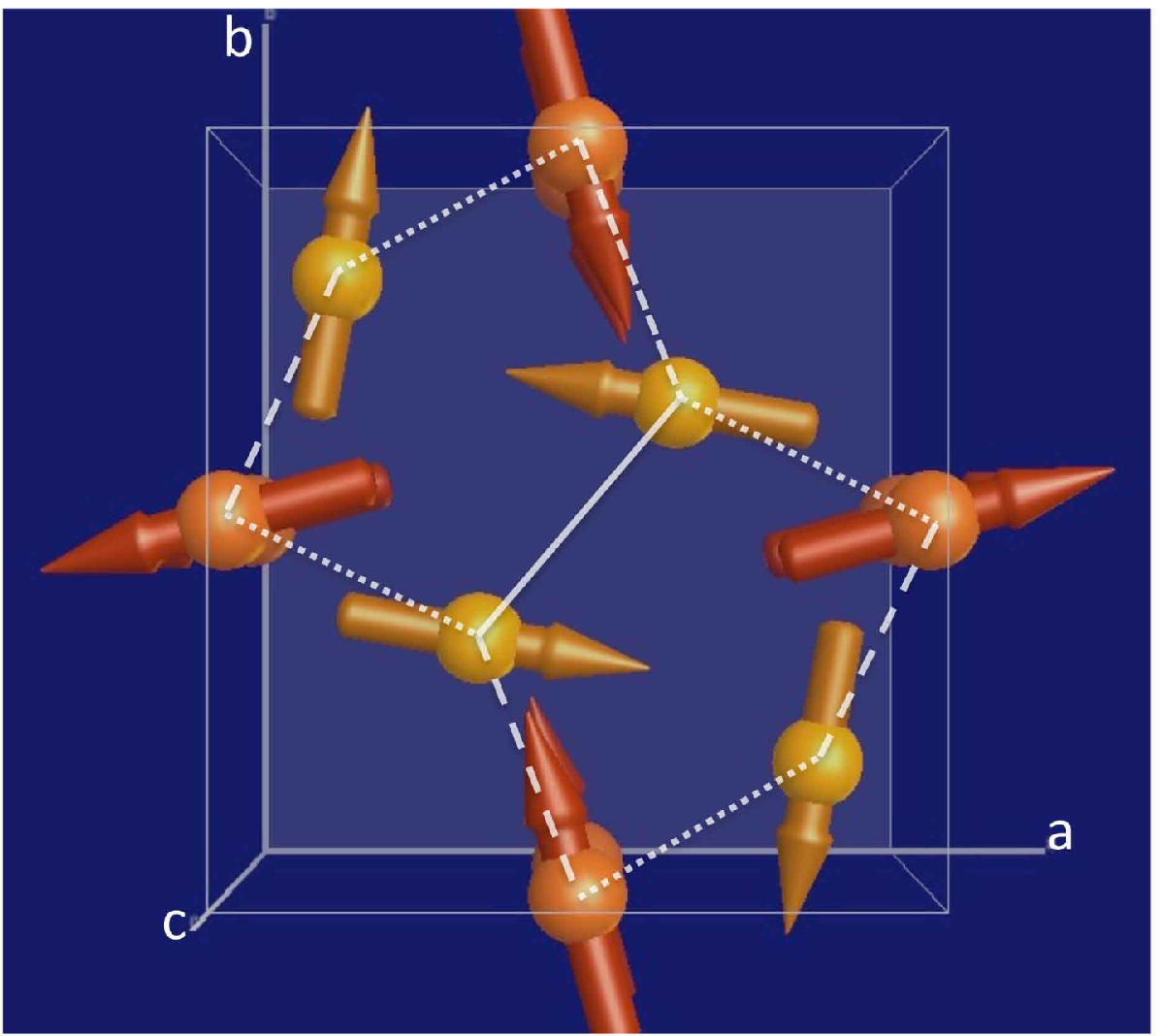}
\caption{(color online) Magnetic structure. First and second magnetic domains. (Yellow): Fe$_1$ sites. (Orange): Fe$_2$ sites. The second domain is deduced from the first one by applying the symmetry operation $(x+1/2, \bar{y}+1/2, \bar{z})$ lost by the magnetic structure, taking into account the translational part that shifts atoms from one cell to a neighbouring one.}
\label{BiFe_magstruc}
\end{figure}

When decreasing the temperature, new reflections, characteristic of an antiferromagnetic order, appear around T$_N = 244(2)$~K. They can be indexed with a propagation vector ${\bf k} = (1/2, 1/2, 1/2)$, that is a doubling of the magnetic cell compared to the nuclear one in the three directions. The temperature variation of  a magnetic Bragg peak is presented in Fig.~\ref{Magn}. The magnetic structure was solved from the 654 magnetic Bragg peaks collected at T$=15$~K taking into account the presence of two magnetic domains \cite{Ressouche}. There are four Fe$_1$ atoms and four Fe$_2$ in the unit cell, whose coordinates are reported in Table~\ref{resneut}. The best refinement led to a residual factor $Rw = 6.4 \%$, with two nearly equally populated domains (42/58\%). The magnetic moment parameters for the first domain are reported in Table~\ref{resneut}. The corresponding pictures of the magnetic arrangements are presented in Fig.~\ref{BiFe_magstruc}.
The moments on all the atoms are found restricted in the $(\bf{a}, \bf{b})$ plane. On each site, the four atoms are gathered into two pairs, antiferromagnetically coupled in the case of Fe$_1$ and ferromagnetically  coupled for Fe$_2$. The magnetic moments of each pair are oriented at $90^\circ$, that is {\bf perpendicular} to the moments of the other pair on the same site. The phase between the two sublattices Fe$_1$ and Fe$_2$ is  $155^\circ$.  The moment amplitudes on the two sites Fe$_1$ and Fe$_2$ are slightly different, and both much smaller than the expected value for a Fe$^{3+}$ ion (5$\mu_B$). A spin transfer process from the iron to the neighbouring oxygen ions \cite{Fe3O4} could explain this reduction.

As it can be seen on Fig.~\ref{BiFe_magstruc}, this non-collinear magnetic structure forms imbricated rectangles of spins. This is quite unusual in such an orthorhombic system, where all the directions are non-equivalent and where there is no fourfold axis. This peculiarity certainly has an origin in the frustration of the magnetic interactions. One should also notice that for this structure the magnetic space group is centrosymmetric, that is non polar. This result  is {\it a priori} not compatible with ferroelectricity, and this  shed some doubts on the claimed ferroelectricity observed in polycrystalline \BiFe~\cite{singh}. 

To check the relevance of the Cairo pentagon lattice model and the role of geometric frustration on the observed magnetic structure of  \BiFe, we have undertaken a two step theoretical approach based on a localized spin model. 
 Calculations have been made both for an ideal Cairo lattice (including the perfect case $J_3 = J_4 = J_5$), and in the real geometry. The angles between magnetic moments in a system is imposed by the set of magnetic interactions. In a first step, an exact Fourier space analysis at zero temperature is performed, assuming a single-k like ordering, which allows to find the periodicity of the magnetic structure \cite{bertaut}. Once the periodicity is known, an energy minimization in real space gives the respective orientation of the magnetic moments within the magnetic unit cell. For these calculations, the five exchange interaction parameters $J_1$ to $J_5$ already described have been considered, using a Heisenberg Hamiltonian $\mathcal{H}=-\frac{1}{2}\sum_{i,j} J_{k}\vec{S_i}\cdot\vec{S_j}$. All the moments were assumed to be of equal amplitudes. 

In the Cairo pentagonal lattice, a large set of $J_i$ values yield the observed (1/2, 1/2, 1/2) propagation vector. The doubling of the magnetic cell along {\bf c} alone requires however $J_2$ to be negative and $J_1$ positive or slightly negative. Whatever the ${\bf k} = (1/2, 1/2, 1/2)$ solution found, the magnetic configuration is always made of antiparallel Fe$_1$ pairs of moments and parallel Fe$_2$ ones. On a same site, the pairs of moments are {\bf perpendicular} to each other.  This is exactly what is observed experimentally. The only parameter that could vary as a function of the  particular values of the $J_i$'s is the phase angle between the two sites Fe${_1}$ and Fe${_2}$. The $\sim 155^{\circ}$ observed value requires $J_3$, $J_4$, $J_5$ to be all negative, with a ratio $J_3$/$J_5$~=2.15 for $\lvert J_4\lvert\geq \lvert J_3\lvert$. The observed magnetic structure of \BiFe\, and in particular the 90$^{\circ}$ phase angle between pairs of atoms at the same site, is thus recovered by this model with some realistic set of the five super-exchange interactions $J_i$. In the \BiFe\ real geometry, the Fe$^{3+}$ lattice differs from the Cairo pentagonal one by the presence of two stacked Fe$_2$ atoms, instead of a single atom, at two vertices in each pentagon. Performing the calculations in the real geometry yield very similar results than those of the ideal case. The perpendicular ferromagnetic and antiferromagnetic Fe$_1$ and Fe$_2$ pairs is a robust recovered feature. The only difference between both lattices is the domain of existence of this phase that extends to lower $\lvert J_3\lvert$ and $\lvert J_5\lvert$ with respect to $\lvert J_4\lvert$ in the real case. The \BiFe\ real lattice is therefore topologically equivalent to the Cairo one after renormalization of the $J_3$ and $J_5$ interactions with respect to the $J_4$ one. 

To summarize,  we have found that the compound \BiFe\  displays a very original non-collinear magnetic structure,  made of  four Fe$_1$ moments and four Fe$_2$ ones forming interpenetrating patterns of four-fold spin rotations. This magnetic arrangement is very peculiar for a pentagonal lattice with nearest neighbour interactions only and is not a mere propagation of the magnetic arrangement minimizing the energy in each individual pentagon, contrary to the case of triangles based lattices. It actually results from two ingredients, a high geometrical frustration and an additional complex connectivity (two kinds of sites with different coordinations). This first materialisation of a pentagonal Cairo lattice opens new perspectives in the field of magnetic frustration.  

\begin{acknowledgments}
 Work at Sofia was supported by the Bulgarian Science Foundation under Grants No. TK-X-1712/2007. The authors are grateful to Michael Mikhov for the high temperature susceptibility results and to Luc Valentin for the help with the tilings.
\end{acknowledgments}

\end{document}